\documentclass[preprint,12pt]{elsarticle}

\usepackage{amsmath,euscript,amssymb,epsfig,amsfonts,latexsym}
\usepackage{graphicx,color}

\newcommand{\be}[1]{\begin{equation}\label{#1}}
\newcommand{\ee}{\end{equation}}
\newcommand{\ba}[1]{\begin{eqnarray}\label{#1}}
\newcommand{\ea}{\end{eqnarray}}
\newcommand{\rf}[1]{(\ref{#1})}
\newcommand{\nn}{\nonumber}

\begin{document}

\begin{frontmatter}

\title{Screening vs. gevolution: \\in chase of a perfect cosmological simulation code}

\author[Maxim]{Maxim Eingorn\corref{mycorrespondingauthor}}
\cortext[mycorrespondingauthor]{Corresponding author}
\ead{maxim.eingorn@gmail.com}

\author[Emrah]{A.~Emrah Y\"{u}kselci}
\author[Alexander]{Alexander Zhuk}

\address[Maxim]{Department of Mathematics and Physics, North Carolina Central University, \\1801 Fayetteville St., Durham, North Carolina 27707, U.S.A.}
\address[Emrah]{Department of Physics, Istanbul Technical University, 34469 Maslak, Istanbul, Turkey}
\address[Alexander]{Astronomical Observatory, Odessa I.I. Mechnikov National University, \\ Dvoryanskaya St. 2, Odessa 65082, Ukraine}

\begin{abstract}
We compare two competing relativistic approaches to the N-body simulation of the Universe large-scale structure. To this end, employing the corresponding alternative computer codes (``gevolution'' and ``screening''), we conduct a series of cosmological simulations in boxes of different sizes and calculate the power spectra of the scalar perturbation $\Phi$, the frame-dragging vector potential ${\bf B}$ and the difference between scalar modes $\chi=\Phi-\Psi$. We demonstrate that the corresponding power spectra are in very good agreement between the compared schemes. For example, the relative difference of the power spectra for $\Phi$ is 0.04\% maximum. Since the perturbed Einstein equations have much simpler form in the “screening” approach, the simulation with this code consumes less computational time, saving almost 40\% of CPU hours.
\end{abstract}


\end{frontmatter}

\

\section{Introduction}
\label{sec:introduction}

The starry sky overhead is one of the most beautiful phenomena on Earth. Therefore, since its inception, humanity has tried to comprehend the nature of this phenomenon. Our ideas changed in the course of technological progress from observations with the naked eye to powerful ground-based and space telescopes. Due to these telescopes, we found that the Universe is filled with a cosmic web which is composed of interconnected filaments of galaxies separated by giant voids \cite{Umehata}. The emergence of this large-scale structure is one of the major challenges of modern cosmology.

The formation of structures with matter density contrast less than unity (linear regime) is well described by analytical methods \cite{Mukhanov2,Rubakov}. However, this task becomes much more difficult in the nonlinear regime, when the density contrast exceeds unity. In this case, the process of structure growth is described employing numerical simulations \cite{Davis}. Gravity is the main force in cosmology responsible for the structure formation \cite{Peebles}. It is known that Newtonian N-body simulations describe well this process in the case of the $\Lambda$CDM (Lambda cold dark matter) model \cite{Springeletal}. However, the Newtonian approach has a number of limitations. Firstly, the Newtonian approximation does not take into account relativistic effects that occur at large cosmological scales. For example, it is insensitive to the existence of a particle horizon. Secondly, it is not applicable for objects with relativistic peculiar velocities. Thirdly, it is problematic to apply the Newtonian approach to theories beyond the $\Lambda$CDM model.
Fourthly, the Newtonian approximation is not appropriate for calculating the effect of backreaction of perturbations on the metric. Although such an effect might be not large, it can nevertheless be measured in the near future. This becomes relevant in the light of upcoming surveys like Euclid, Square Kilometre Array and Large Synoptic Survey Telescope \cite{Euclid,SKA,LSST}. Therefore, it seems very desirable to formulate a theory that eliminates the shortcomings of Newtonian cosmology and works at all scales, from relatively small astrophysical ones (although not in immediate proximity to black holes or neutron stars) to largest cosmological distances. Hence, such a theory must be valid for both the nonlinear and linear regimes. It is clear that this goal can be achieved in the framework of general relativity. Since the gravitational field is weak at the considered scales, one can apply the theory of perturbations.

The corresponding scheme was developed in \cite{Adamek1,Adamek2}, serving as the foundation for the N-body cosmological simulation code ``gevolution'' \cite{gevNat,gevolution}. This is a relativistic code based on the weak field expansion. Along with the gravitational potential $\Psi$, which is also determined within the Newtonian approximation, the code makes it possible to determine the non-Newtonian degrees of freedom of the metric: the scalar one $\Phi-\Psi$, the frame-dragging vector potential ${\bf B}$, and two helicities $h_{ij}$ of gravitational waves. A characteristic feature of this scheme is that the equations include not only linear terms, but also those which are quadratic in scalar perturbations. As a result, metric corrections represent mixtures of the first- and second-order quantities.

Mixing of orders of smallness leads to a rather complicated form of equations. Therefore, it seems logical to construct a perturbation theory where orders of smallness are not mixed. In this case, the lowest-order metric perturbations are determined first, and then they serve as sources for the higher-order quantities. Such a theory was proposed in \cite{Eingorn1,BrilEin,Duygu}. Here, the gravitational field is weak and peculiar velocities are much less than the speed of light. These are the only limitations of this scheme. All equations are linear that allows to solve them analytically in the case of the $\Lambda$CDM model \cite{Eingorn1,Duygu}. A characteristic feature of this approach is that the equation for the first-order gravitational potential has the form of the Helmholtz equation, not the Poisson one. As a result, its solution has the form of the Yukawa potential with an exponential cutoff at large cosmological distances \cite{Eingorn1,MaxEzgi}. This ``screening'' scheme was generalized in \cite{fluids1,fluids2,fluids3,fluids4} to the cases of models with perfect fluids, both with linear and nonlinear equations of state.

Since we have two competing relativistic approaches, it is very interesting to compare them with respect to the N-body simulation of the large-scale structure formation. This is the main goal of the present paper. With the help of the corresponding alternative computer codes, we calculate the power spectra of $\Phi$, ${\bf B}$ and $\Phi-\Psi$ for different box sizes. One should keep in mind that we use the same notation for these degrees of freedom, although the ``gevolution'' quantities $\Phi$ and ${\bf B}$ have the second-order admixtures. Nevertheless, we demonstrate that the corresponding power spectra are in very good agreement between the compared schemes. Hence, the effect of the second-order admixtures is indeed small. For example, the relative difference of the power spectra for $\Phi$ is 0.04\% maximum. On the other hand, employing the simpler ``screening'' code saves almost 40\% of the expensive computational time.  Additionally, for the fixed allotted time, the faster code makes it possible to simulate a larger box or to probe higher resolution, which is a definite advantage. 

The paper is structured as follows. In Section 2, we briefly describe the ``gevolution'' scheme. Section 3 is devoted to the ``screening'' one. In Section 4, we present the simulation outcomes for both of these approaches, namely, the power spectra of the metric perturbations, their relative difference, and the computational time consumption. The main results are summarized in concluding Section 5.

\

\section{gevolution}
\label{sec:gevolution}

This section is devoted to the equations underlying simulations by means of the original ``gevolution'' code \cite{gevNat,gevolution}. First of all, let us write down the perturbed Friedmann-Lema\^{\i}tre-Robertson-Walker metric
\be{1}
ds^2 = a^2 \left[ -(1+2\Psi)d\tau^2 - 2 B_i dx^i d\tau + (1-2\Phi)\delta_{ij} dx^i dx^j \right]\, ,\quad i,j=1,2,3\, ,
\ee
in the Poisson gauge. Here $a(\tau)$ is the scale factor (depending on the conformal time $\tau$); $x^i$ denote the comoving coordinates; $\Psi$ and $\Phi$ represent the scalar perturbations, while $B_i$ constitute the vector perturbation and satisfy the gauge condition $\delta^{ij} B_{i,j}=0$ (where $_{,j}$ stands for $\partial/\partial x^j$). Throughout the narration, tensor modes are disregarded.

If one demands the smallness of metric corrections $\Psi$, $\Phi$, $B_i$ (the weak gravitational field regime) and keeps those terms which are linear in them, along with the terms quadratic in $\Psi$, $\Phi$ which contain two spatial derivatives (see the summands proportional to $\Phi\Delta\Phi$, $\Phi_{,i}\Phi_{,j}$, $\chi\Phi_{,ij}$ and $\Phi\Phi_{,ij}$ below), then the Einstein equations are reduced to the following ones (see Eqs.~(2.9), (2.11) and (2.10) in \cite{gevolution}, respectively):
\be{2}
(1 + 4\Phi) \Delta\Phi - 3\mathcal{H}\Phi' + 3\mathcal{H}^2(\chi - \Phi) + \frac{3}{2} \delta^{ij} \Phi_{,i} \Phi_{,j} = - 4\pi G a^2 \left( T_0^0 - \overline{T_0^0} \right)\, ,
\ee
\be{3}
-\frac{1}{4} \Delta B_i - \Phi'_{,i} - \mathcal{H}(\Phi_{,i} - \chi_{,i}) = 4\pi G a^2 T_i^0\, ,
\ee
\ba{4}
&&\left( \delta_k^i \delta_l^j - \frac{1}{3} \delta^{ij} \delta_{kl} \right) \left[ B'_{(i,j)} + 2\mathcal{H} B_{(i,j)} + \chi_{,ij} - 2\chi\Phi_{,ij} + 2\Phi_{,i} \Phi_{,j} + 4\Phi\Phi_{,ij} \right]\nn\\
&=&8\pi G a^2 \left( \delta_{ik} T^i_l - \frac{1}{3} \delta_{kl} T_i^i \right)\, ,
\ea
where $\Delta\equiv \delta^{ij}\partial^2/\partial x^i\partial x^j$ is the Laplacian in comoving coordinates; $\mathcal{H}\equiv a'/a$, with prime denoting the derivative with respect to $\tau$; $\chi\equiv\Phi-\Psi$ represents the difference between scalar modes; $G$ is the gravitational constant; finally, $_{(ij)}$ stands for the symmetrization over indexes $i,j$. We regard Eqs.~\rf{2}, \rf{3} and \rf{4} as equations for $\Phi$, $B_i$ and $\chi$, respectively. Components of the energy-momentum tensor in their right-hand sides follow from corresponding Eqs.~(3.7), (3.10) and (3.8) in \cite{gevolution}, when Eq.~(3.11) therein is taken into account: 
\be{5}
T^0_0 = - \frac{1}{a^4} \sum_n \delta({\bf r}-{\bf r}_n) \sqrt{q^2_n + m^2_n a^2} \left( 1 + 3\Phi + \frac{q^2_n}{q^2_n + m^2_n a^2} \Phi \right)\, ,
\ee
\be{6}
T^0_i = \frac{1}{a^4} \sum_n \delta({\bf r}-{\bf r}_n) (q_n)_i ( 1 + 2\Phi + \chi)\, ,
\ee
\be{7}
T^i_j = \frac{\delta^{ik}}{a^4} \sum_n \delta({\bf r}-{\bf r}_n) \frac{(q_n)_j  (q_n)_k}{\sqrt{q^2_n + m^2_n a^2}} \left( 1 + 4\Phi + \frac{m^2_n a^2}{q^2_n + m^2_n a^2} \Phi \right)\, ,
\ee
where $q_n^2\equiv\delta^{ij}(q_n)_i(q_n)_j$. These components describe a system of point-like particles with masses $m_n$, comoving radius-vectors ${\bf r}_n$, and momenta ${\bf q}_n$, which satisfy the equation of motion (see Eq.~(3.5) in \cite{gevolution})
\be{8}
\frac{d(q_n)_i}{d\tau} = - \sqrt{q_n^2 + m_n^2 a^2} \left[ \left( 1 + \frac{q_n^2}{q_n^2 + m_n^2 a^2} \right) \Phi_{,i} - \chi_{,i} + \frac{\delta^{jk}(q_n)_k B_{j,i}}{\sqrt{q_n^2 + m_n^2 a^2}} \right]\, .
\ee

It is very important to emphasize that the presented formalism covers the whole space (except for its small portion where gravity is strong) as the smallness of energy-momentum fluctuations (generating the inhomogeneous gravitational field) is not demanded \cite{gevNat,gevolution}. In particular, there exist spatial regions where the absolute value of $T_0^0$ is much larger than that of the background quantity $\overline{T_0^0}$. The latter enters into the Friedmann equation for the scale factor $a(\tau)$:
\be{9}
-\frac{3\mathcal{H}^2}{a^2} = 8\pi G \left(\overline{T^0_0}+\ldots\right)\, ,
\ee
where marks of omission $\ldots$ comprise additional contributions from radiation (treated as homogeneous) and the cosmological constant (in the case of concordance $\Lambda$CDM model).

\

\section{Screening}
\label{sec:screening}

In this section we simplify the aforesaid equations in the spirit of the all-scale cosmological perturbation theory formulated in \cite{Eingorn1,BrilEin} (subsequently also referred to as the cosmic screening approach \cite{Duygu,MaxEzgi} due to its prediction of Yukawa interparticle interaction). The corresponding modified $N$-body code we call ``screening''.

Similarly to \cite{gevNat,gevolution}, the first-order formalism \cite{Eingorn1} is valid for all distances, provided that the metric corrections $\Psi$, $\Phi$, $B_i$ are small in every place of interest (unlike the energy-momentum fluctuations which may be large). However, these corrections are strictly assigned the first order of smallness, and so the terms quadratic in them are left out as belonging to the second-order theory, irrespective of the presence of spatial derivatives (see \cite{BrilEin} for compelling evidence of self-consistency of such assignments). Furthermore, since nonrelativistic matter is in the focus of attention, only terms which are linear in particles' momenta (and do not contain metric perturbations) are kept, 
resulting in the coincidence $\Psi=\Phi$ and automatically turning $\chi$ into a second-order quantity. Then, instead of Eqs.~\rf{2} and \rf{3}, the Einstein equations for $\Phi$ and $B_i$ take the form
\be{10}
\Delta\Phi - 3\mathcal{H}\Phi' - 3\mathcal{H}^2\Phi = - 4\pi G a^2 \left( T_0^0 - \overline{T_0^0} \right)\, ,
\ee
\be{11}
-\frac{1}{4} \Delta B_i - \Phi'_{,i} - \mathcal{H}\Phi_{,i} = 4\pi G a^2 T_i^0\, ,
\ee
where, replacing Eqs.~\rf{5} and \rf{6},
\be{12}
T^0_0 = - \frac{1}{a^3} \sum_n m_n\delta({\bf r}-{\bf r}_n) \left( 1 + 3\Phi \right)\, ,
\ee
\be{13}
T^0_i = \frac{1}{a^4} \sum_n \delta({\bf r}-{\bf r}_n) (q_n)_i \, .
\ee
Moreover, introducing the mass density
\be{14}
\rho=\sum_n m_n\delta({\bf r}-{\bf r}_n)\, ,
\ee
along with its background value $\overline\rho$ and the corresponding fluctuation $\delta\rho\equiv\rho-\overline\rho$, from \rf{12} we derive (see Eq.~(2.13) in \cite{Eingorn1})
\be{15}
T^0_0 - \overline{T^0_0} = - \frac{1}{a^3} \delta\rho - \frac{3}{a^3} \overline\rho \Phi\, .
\ee
Here $\overline{T^0_0}=-\overline\rho/a^3$, and $\rho$ in the term $\propto\rho\Phi$ is replaced by $\overline\rho$ in view of the fact that $|\delta\rho\Phi|\ll|\delta\rho|$ for $|\Phi|\ll1$ (the omitted term $\propto\delta\rho\Phi$ generates the second-order perturbations \cite{BrilEin}). Substituting \rf{15} into \rf{10}, one easily recognizes Eq.~(2.15) in \cite{Eingorn1}. Obviously, this linear equation for $\Phi$ is much simpler than nonlinear Eq.~\rf{2}. It is worth mentioning that Eq.~\rf{10} with the right-hand side proportional to \rf{15} admits the exact analytical solution \cite{Eingorn1}. 

Finally, the equation of motion \rf{8} is noticeably simplified as well:
\be{16}
\frac{d(q_n)_i}{d\tau} = - m_n a \Phi_{,i} \, .
\ee

Now, if we would like to take into account the summands quadratic in $\Phi$, $B_i$, $(q_n)_i$ and $\delta\rho$ in order to investigate such nonlinear effects as, for instance, cosmological backreaction, then the second-order formalism \cite{BrilEin} should be employed. In its framework, such summands play the role of sources for second-order metric corrections, which can be easily determined from the corresponding bulky, but linear equations. When further simplified, these equations can be even solved analytically \cite{Duygu}. In particular, the metric coefficients $g_{00}$ and $g_{11,22,33}$ acquire the second-order scalar perturbations $-2a^2\Psi^{(2)}$ and $-2a^2\Phi^{(2)}$, respectively. Hence, now $\chi\equiv\Phi^{(2)}-\Psi^{(2)}$. The direct way to estimate this difference between scalar modes with the help of the existing algorithm is to employ Eq.~(C.9) from \cite{gevolution}, substituting there $\Phi$ determined from \rf{10} and neglecting the higher-order terms such as those containing simultaneously both $\chi$ and $\Phi$. Instead of computing $\chi$ at each iteration step, one can calculate this and other second-order quantities only at redshifts of interest, thereby saving valuable computational time (which may be already substantially reduced owing to modification of equations).

It is important to keep in mind that we use the same letter for the function $\Phi$ from the previous section (``gevolution'') and for the first-order function $\Phi$ from the current section (``screening''), though actually $\Phi_{\mathrm{gevolution}}$ equals $\Phi_{\mathrm{screening}}$ plus a higher-order admixture. A similar remark applies to $B_i$ too. Generally, from the point of view of \cite{Eingorn1,BrilEin}, the first and second orders of perturbations are mixed in \cite{gevNat,gevolution}, and this seems an unnecessary complication, provided that they can be successfully separated.

\

\section{Simulations}
\label{sec:simulations}

\subsection{Inputs}

A couple of relevant questions arises immediately. First, do the ``gevolution'' and ``screening'' codes produce different or almost identical results? Second, which code runs faster?

Looking for the answers, we have conducted a series of cosmological N-body simulations in boxes of sizes $280$, $336$, $560$, $980$, $1680$~Mpc/h with $1$~Mpc/h resolution as well as an additional series in boxes of sizes $280$, $560$, $1120$, $2016$, $2800$~Mpc/h with $2$~Mpc/h resolution. The specified comoving dimensions $L$ of cubic boxes coincide with the physical ones $L_{\mathrm{ph}}=L/(1+z)$ at the present time (when the redshift $z$ is $0$). The initial redshift has been set to $100$ for all simulations.

Regarding the Universe ingredients, baryons have been treated similarly to cold dark matter. In addition, 
radiation has been neglected in both codes, with no effect on the conclusions.

In order to generate initial conditions, we have used the code CLASS \cite{CLASS}. Finally, the following values of standard cosmological parameters have been chosen \cite{Planck2018}: $\Omega_b h^2=0.02242$, $\Omega_c h^2=0.11933$ (where $h=0.6766$), $n_s=0.9665$, $A_s=2.105\times 10^{-9}$, $k_{\mathrm{pivot}}=0.05$~Mpc$^{-1}$.

\

\subsection{Outputs}

\begin{figure*}
	\resizebox{1\textwidth}{!}{\includegraphics{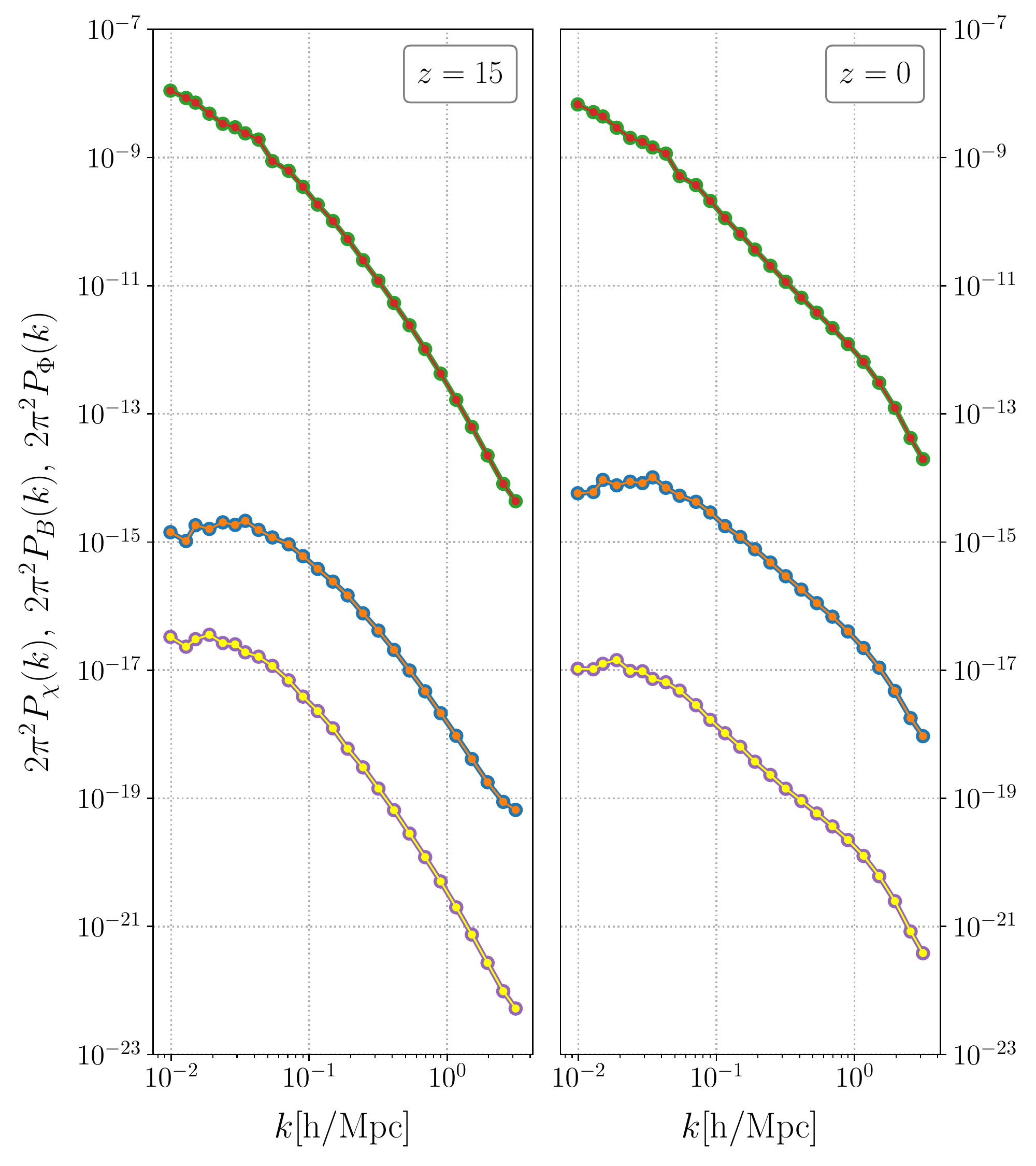}}
	\caption{Power spectra of $\Phi$ ({\bf top} curves), ${\bf B}$ ({\bf middle} curves) and $\chi$ ({\bf bottom} curves) from the ``gevolution'' code (green, blue and purple curves in the background, respectively) and from the ``screening'' code (red, orange and yellow curves in the foreground, respectively) at redshifts $z=15$ ({\bf left} graph) and $z=0$ ({\bf right} graph). The simulation box size amounts to $980$~Mpc/h.}
	\label{fig:1}
\end{figure*}

\begin{figure*}
	\resizebox{1\textwidth}{!}{\includegraphics{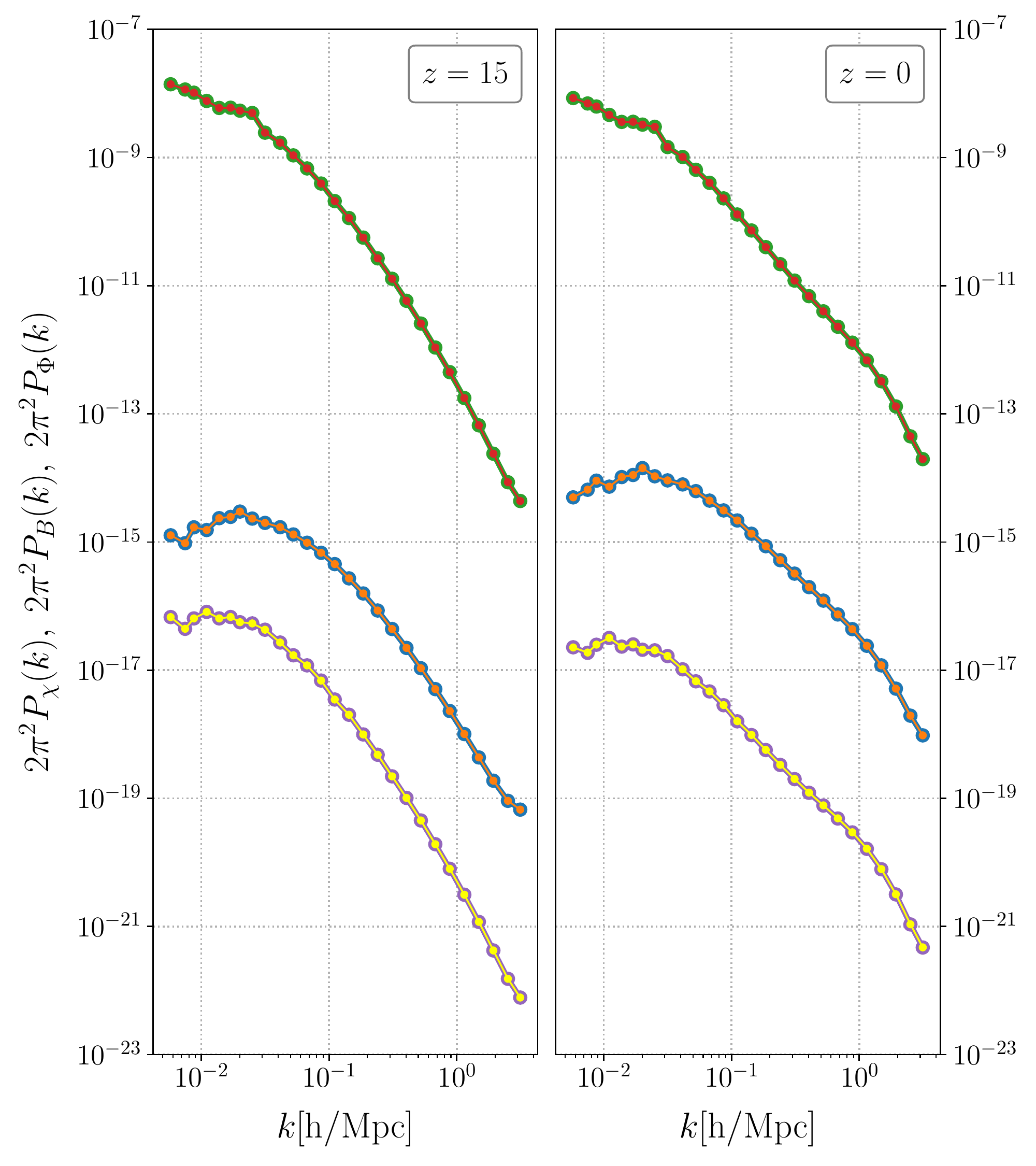}}
	\caption{Power spectra of $\Phi$ ({\bf top} curves), ${\bf B}$ ({\bf middle} curves) and $\chi$ ({\bf bottom} curves) from the ``gevolution'' code (green, blue and purple curves in the background, respectively) and from the ``screening'' code (red, orange and yellow curves in the foreground, respectively) at redshifts $z=15$ ({\bf left} graph) and $z=0$ ({\bf right} graph). The simulation box size amounts to $1680$~Mpc/h.}
	\label{fig:2}
\end{figure*}

\begin{figure*}
	\resizebox{1\textwidth}{!}{\includegraphics{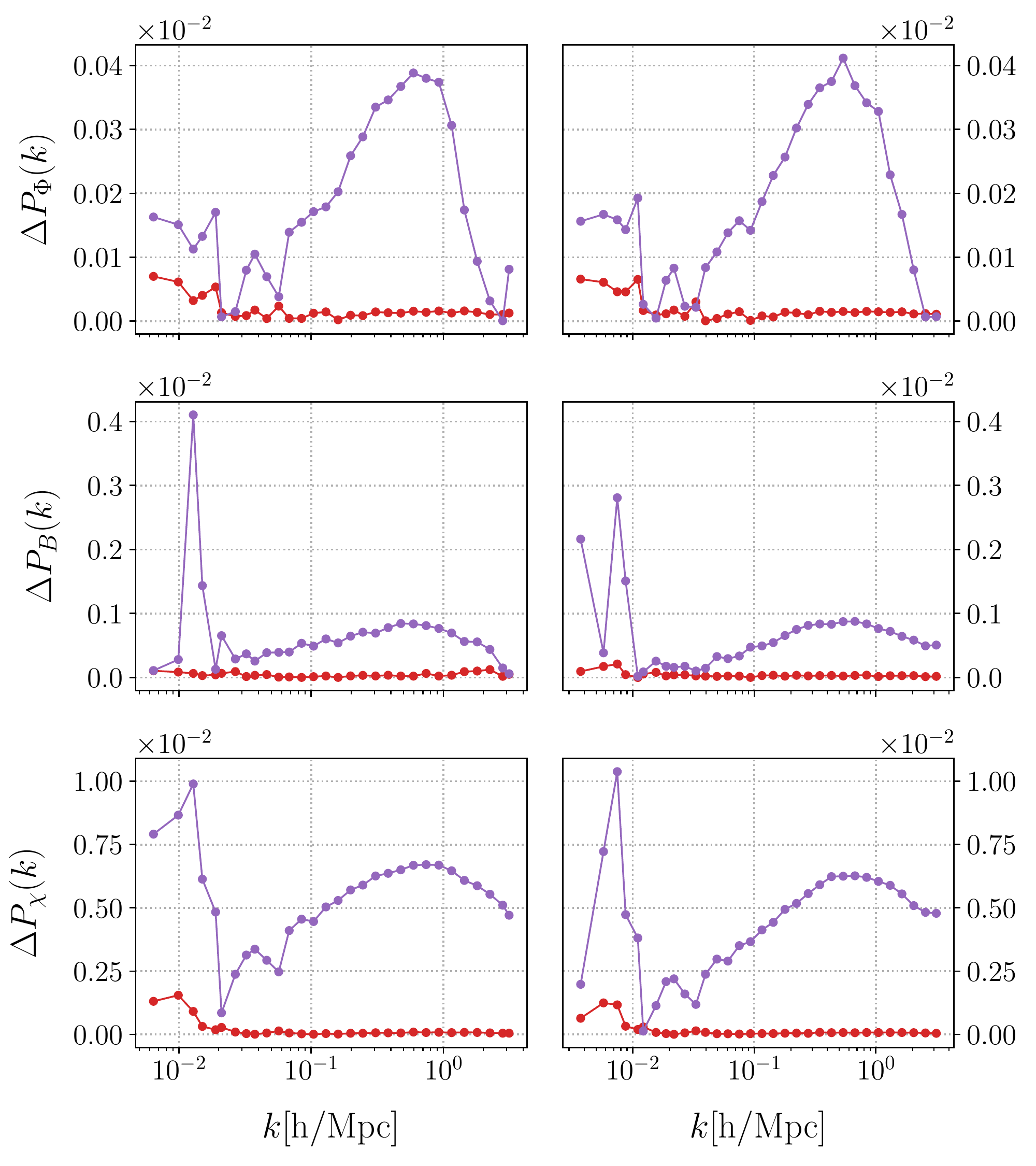}}
	\caption{Relative deviations of the power spectra of $\Phi$ ({\bf top} graphs), ${\bf B}$ ({\bf middle} graphs) and $\chi$ ({\bf bottom} graphs) predicted by the ``screening'' code from the ``gevolution'' code counterparts at redshifts $z=15$ (red) and $z=0$ (purple). The simulation box size amounts to $980$~Mpc/h for all three {\bf left} graphs and $1680$~Mpc/h for all three {\bf right} graphs.}
	\label{fig:3}
\end{figure*}

The main outcomes of our simulations by both the ``gevolution'' and ``screening'' codes are the power spectra of $\Phi$, $B_i$ and $\chi$. These power spectra ($P_{\Phi}$, $P_{B}$ and $P_{\chi}$, respectively) are defined by the formulas \cite{gevNat}
\be{17}
4\pi k^3 \left\langle \Phi({\bf k},z)\Phi({\bf k}',z) \right\rangle = (2\pi)^3\delta({\bf k}-{\bf k}')P_{\Phi}(k,z)\, ,
\ee
\be{18}
4\pi k^3 \left\langle B_i({\bf k},z)B_j({\bf k}',z) \right\rangle = (2\pi)^3\delta({\bf k}-{\bf k}')P_{ij}P_{B}(k,z)\, ,\quad P_{ij}\equiv\delta_{ij}-\frac{k_ik_j}{k^2}\, ,
\ee
\be{19}
4\pi k^3 \left\langle \chi({\bf k},z)\chi({\bf k}',z) \right\rangle = (2\pi)^3\delta({\bf k}-{\bf k}')P_{\chi}(k,z)\, .
\ee
To estimate quantitatively the difference between two analyzed competing approaches, we introduce the relative deviation
\be{20}
\Delta P\equiv \left|\frac{P_{\mathrm{screening}}-P_{\mathrm{gevolution}}}{P_{\mathrm{gevolution}}}\right|
\ee
for each power spectrum. In Figures~\ref{fig:1}-\ref{fig:3} we depict the sought power spectra and relative deviations for two boxes with comoving sizes $980$~Mpc/h and $1680$~Mpc/h, at redshifts $z=15$ and $z=0$. For the former box, $k$ ranges from $2\pi/980\approx6.4\times10^{-3}$~h/Mpc to $\pi$~h/Mpc, while for the latter one, $k$ ranges from $2\pi/1680\approx3.7\times10^{-3}$~h/Mpc again to $\pi$~h/Mpc. 
These figures demonstrate that the results of ``gevolution'' and ``screening'' simulations remarkably coincide with each other. For example, the maxima of relative deviations at the present time ($z=0$) are approximately 0.04\%, 0.4\% and 1\% for $P_{\Phi}$, $P_B$ and $P_{\chi}$, respectively. At redshift $z=15$, the corresponding relative deviations are much less than these values. 

\ 

\ 

\subsection{Computational time consumption}

The resource consumption time is an important parameter of an N-body simulation: the smaller this time, the cheaper the project costs. This is especially important for projects running on supercomputers. It is also clear that the faster code can simulate a larger box for the allotted time. Therefore, we compare the computational time (in CPU hours) for the ``gevolution'' ($t_{\mathrm{g}}$) and ``screening'' ($t_{\mathrm{s}}$) approaches by estimating the relative deviation
\be{21}
\Delta t\equiv \left|\frac{t_{\mathrm{s}}-t_{\mathrm{g}}}{t_{\mathrm{g}}}\right|\, .
\ee
The results of such a comparison are presented in Table~\ref{table:1} for different simulation box sizes and for two resolutions. As one can see from this table, in all cases the ``screening'' code runs approximately 40\% faster than the rival.

\ 

\begin{table}[h!]
	\centering
	\begin{tabular}{c|c|c|c}
		$L\,$[Mpc/h] & $t_{\mathrm{g}}$ & $t_{\mathrm{s}}$ & $\Delta t\,$[\%] \\[0.1cm]
		\hline
				
		280 & 6.2 & 3.8 & 38.7 \\
		
		336 & 11.5 & 7.1 & 38.3 \\
		
		560 & 47.1 & 29.2 & 38.0 \\
		
		980 & 294.9 & 180.9 & 38.7 \\
		
		1680 & 1551.1 & 939.6 & 39.4 \\
		
	\end{tabular} \ \ \begin{tabular}{c|c|c|c}
		$L\,$[Mpc/h] & $t_{\mathrm{g}}$ & $t_{\mathrm{s}}$ & $\Delta t\,$[\%] \\[0.1cm]
		\hline
		
		280 & 0.8 & 0.5 & 37.5 \\
		
		560 & 4.3 & 2.7 & 37.2 \\
		
		1120 & 34.8 & 21.2 & 39.1 \\
		
		2016 & 216.2 & 132.1 & 38.9 \\
		
		2800 & 563.9 & 347.0 & 38.5 \\
		
	\end{tabular}
	
	\caption{\label{table:1} Computational time (in CPU hours) for the ``gevolution'' ($t_{\mathrm{g}}$) and ``screening'' ($t_{\mathrm{s}}$) approaches and the relative deviation $\Delta t$ (in \%) for different simulation box sizes $L$ and for two resolutions: $1$~Mpc/h ({\bf left} chart) and $2$~Mpc/h ({\bf right} chart).}
\end{table}

\ 

\section{Conclusion}
\label{sec:conclusion}

In the present paper we have considered two alternative relativistic approaches to the N-body simulation of the Universe large-scale structure. The first one \cite{Adamek1,Adamek2} is the basis for the simulation code ``gevolution'' \cite{gevNat,gevolution}. The main features of this approach are the following ones: first, it determines non-Newtonian degrees of freedom, second, it works at all scales (from relatively small astrophysical ones to largest cosmological distances), third, it gives an opportunity to take into account relativistic particles. However, since the first and second orders of smallness are mixed, the corresponding perturbed Einstein equations have rather complicated form. These equations are greatly simplified if orders of smallness are clearly separated from each other. Such a scheme was proposed in \cite{Eingorn1,BrilEin,Duygu}. Within this scheme, the first-order corrections for metric coefficients are determined first, and then they serve as sources for the second-order ones. All resulting equations are linear, which allows them to be solved analytically in the case of the $\Lambda$CDM model \cite{Eingorn1,Duygu}. The simpler structure of equations makes it possible to clarify the physical properties of perturbations. For example, the equation for the first-order gravitational potential has the form of the Helmholtz equation. As a result, its solution has the form of the Yukawa potential with an exponential cutoff at large cosmological distances \cite{Eingorn1,MaxEzgi}. The only limitations of this ``cosmic screening'' approach \cite{Duygu,MaxEzgi} are as follows: the gravitational field is weak (similarly to ``gevolution'') and peculiar velocities are much less than the speed of light. It should be noted that the second condition is alleviated due to the fact that this formalism allows one to take into account not only those summands in the perturbed Einstein equations, which are linear in velocities of particles, but also the quadratic ones \cite{BrilEin}.

The main goal of the present paper was to compare these two competing relativistic schemes with respect to the N-body simulation of the large-scale structure formation. To this end, we have conducted a series of simulations in boxes of sizes 280, 336, 560, 980, 1680 Mpc/h with
1 Mpc/h resolution as well as an additional series in boxes of sizes 280, 560, 1120, 2016, 2800 Mpc/h with 2 Mpc/h resolution. Employing the corresponding alternative computer codes, we have calculated the power spectra of the scalar perturbation $\Phi$, the frame-dragging vector potential ${\bf B}$ and the quantity $\chi=\Phi-\Psi$ within each of the considered approaches. Then, we have determined the relative deviations of the corresponding power spectra. Despite the fact that the ``gevolution'' quantities $\Phi$ and ${\bf B}$ have the second-order admixtures, we have demonstrated that the power spectra are in very good agreement between the compared schemes. For example, the relative difference of the power spectra for $\Phi$ is 0.04\% maximum. Hence, the effect of the second-order admixtures is small, as it should be. 

It is natural to expect that the code using simpler equations consumes less computational time. Indeed, we have shown that the simpler ``screening'' code saves almost 40\% of CPU hours. 
Since the smaller the computational time, the cheaper the project costs, this is a definite advantage of the “screening” approach. Additionally, the faster ``screening'' code can simulate a larger box for the fixed allotted time. 

Thus, our study shows that the use of the screening formalism is quite justified and can be successful for investigation of various problems in the N-body simulation of the Universe large-scale structure.

\ 

\section*{Acknowledgments}

{\noindent Computing resources used in this work were provided by the National Center for High Performance Computing of Turkey (UHeM) under grant number 4007162019.}

\ 


\end{document}